\documentclass[12pt]{iopart}

\usepackage{graphicx} 
\usepackage{subfigure} 
\usepackage{epstopdf}
\usepackage{amssymb}
\usepackage{amsbsy}

\newcommand{\gso}{Gd$_{2}$Sn$_{2}$O$_{7}$}

\begin{document}
\title[Zero Field Muon Relaxation Rate in the Rare Earth Pyrochlore Gd$_2$Sn$_2$O$_7$]
{Calculation of the Expected Zero Field Muon Relaxation Rate in the Geometrically 
Frustrated Rare Earth Pyrochlore Gd$_2$Sn$_2$O$_7$ Antiferromagnet}
\author{P A McClarty$^1$, J N Cosman$^2$, A G Del Maestro$^3$ and M J P Gingras$^{1,4}$}
\address{$^1$ Department of Physics and Astronomy, University of Waterloo, Waterloo, ON, N2L 3G1, Canada.}
\address{$^2$ Department of Economics, University of British Columbia, 997 - 1873 East Mall, Vancouver, BC, V6T 1Z1, Canada}
\address{$^3$ Department of Physics and Astronomy, Johns Hopkins University, 366 Bloomberg Center, 3400 N. Charles Street, Baltimore, MD 21218, USA} 
\address{$^4$ Canadian Institute for Advanced Research, 180 Dundas Street West, Suite 1400, Toronto, ON, M5G 1Z8, Canada.}

\begin{abstract}
The magnetic insulator \gso\ is one of many geometrically frustrated magnetic materials known 
to exhibit a nonzero muon spin polarization relaxation rate, $\lambda(T)$, down to the lowest 
temperature ($T$) studied. Such behaviour is typically interpreted as a significant level
 of persisting spin dynamics (PSD) of the host material. In the case of \gso, such PSD comes
 as a surprise since magnetic specific heat measurements suggest conventional gapped magnons,
 which would naively lead to an exponentially vanishing $\lambda(T)$ as $T \rightarrow 0$. 
In contrast to most 
materials that display PSD, the ordered phase of \gso\ is well characterized
 and both the nature and the magnitude of the interactions have been inferred from 
the magnetic structure and the temperature dependence of the magnetic specific heat. 
Based on this understanding, the temperature dependence of the muon spin polarization 
relaxation through the scattering of spin waves (magnons) is calculated. 
The result explicitly shows that, despite the unusual extensive number of weakly dispersive (gapped) 
excitations characterizing \gso, a remnant of the zero modes of the parent frustrated pyrochlore 
Heisenberg antiferromagnet, the temperature dependence of the calculated $\lambda(T)$ differs
 dramatically from the experimental one. Indeed, the calculation conforms to the naive expectation
 of an exponential collapse of $\lambda(T)$ at temperatures below $\sim 0.7$ K. 
This result, for the first time, illustrates crisply and quantitatively the paradox that
 presents itself with the pervasive occurrence of PSD in highly frustrated
 magnetic systems as evinced by muon spin relaxation measurements.
\end{abstract}

\section{Introduction}

Muon spin rotation ($\mu$SR) \cite{MuonReview,MuonReview2} 
is a local probe that is sensitive to the magnetic environment 
at the microscopic level in condensed matter. Briefly, 
$\mu SR$ experiments involve injecting positive muons 
polarized along some direction into a condensed matter sample. Each muon comes to rest within the sample 
and its spin precesses in the local magnetic environment. It may also be subject to processes that
 cause it to depolarize. The muon is an unstable particle and decays on a time scale of the order of $\mu$s.
 Since it is mediated by weak interactions, the decay cross section in the muon rest frame is anisotropic in space 
with the anisotropy axis depending on the muon polarization direction at the time of the decay. 
One therefore measures the asymmetry of the decay with respect to the initial polarization axis 
to obtain information about the effect of the material on the magnetic moment of the muon.

One of the many uses of $\mu$SR is in probing static and dynamic processes in magnetic 
materials \cite{MuonReview,MuonReview2}. Consider, for example, a conventional insulating
 magnetic material that undergoes a transition to long range order at some temperature $T_{c}$. 
For temperature $T>T_{c}$ and in zero applied magnetic field, the material is a paramagnet with no static moments.
 Injected muons therefore experience only fluctuating fields which bring about an exponential 
relaxation of the asymmetry. In the critical region near $T_c$, the muon relaxation rate increases, 
diverging at $T_c$ as the spins exhibit critical slowing down. 
Below $T_{c}$, muons precess in the local magnetic field. If the magnetic environment is homogeneous, 
the asymmetry exhibits oscillations at the Larmor frequency of the muon. However, samples are often 
inhomogeneous and there is also generally a distribution of muon stopping sites, so the oscillations 
decay or are altogether undetected because 
the asymmetry probes the precessional dynamics averaged over the whole distribution. 
Such relaxation due to static fields, leaves a residual asymmetry which can decay via genuine dynamical 
processes in the sample. In conventional magnets, with long range order or with frozen moments,
 as in spin glasses, this dynamical relaxation rate $\lambda$ should tend to zero as the temperature vanishes. 

In contrast, $\mu$SR studies have uncovered, over the last two decades, many materials in which there 
is anomalous relaxation rate at low temperatures that is temperature independent. Among them are quasi 
two dimensional \cite{Bert,Fudamoto,Keren3,Zorko,Uemura,Keren2} and 3D magnets \cite{Lago}-\cite{LagoSpinIce}.
 Persistent spin dynamics (PSD) has been detected both in long range ordered magnetic 
systems \cite{Lago,Dunsiger1,Yaouanc,deReotier,Bert_TSO,Giblin,Bonville,Chapuis,Kalvius}, 
in candidate spin liquids \cite{Gardner,Keren,Hodges}
 and systems with freezing transitions with no apparent long range order \cite{Hodges,Dunsiger2,Dunsiger,Marshall}.
 Also, there is persistent dynamics in the low temperature spin ice state of the Ising magnet Dy$_{2}$Ti$_{2}$O$_{7}$
 although the temperature dependence has not yet been explored \cite{Harris,LagoSpinIce}. 

To date, there is no quantitative
microscopic theory that accounts for PSD in any material in which the phenomenon 
is seen. The problem is especially paradoxical in magnets with long range order because spin dynamics
 should be frozen out as the temperature is lowered well below the phase transition temperature.
 In this article, we highlight 
the problem by focussing on a geometrically frustrated magnetic material for which the magnetic structure 
has been characterized, the principal interactions identified and for which the low energy excitations 
can be computed reliably. The material in question is the rare earth pyrochlore 
Gd$_{2}$Sn$_{2}$O$_{7}$ \cite{Bonville,Chapuis,Quilliam, Stewart,Sosin,Bertin,Wills}, another material 
that exhibits PSD \cite{Bonville,Chapuis,Bertin}. 
In the next section, we introduce this material in some more detail with an outline of its phenomenology 
and a description of the available microscopic theory for \gso. Section~\ref{sec:statics} is a discussion 
of the field distribution in \gso\ and an estimate of the muon 
position and distribution of fields that the muons experience. 
In Section~\ref{sec:dynamics}, 
we take the spectrum of excitations and compute from it the relaxation rate one would expect below $T_{c}$ 
and compare with the experimental rate, finding significant discrepancy. 
As far as we are aware, this is the first detailed microscopic calculation of the muon spin depolarization 
rate, $\lambda$, in a material exhibiting PSD. Finally, we present in Section~\ref{sec:disc} a comparison
 of our results with the experimental data and close with a general discussion of PSD in
magnetic rare earth pyrochlore oxides.

\section{Gd$_{2}$Sn$_{2}$O$_{7}$}

\gso\ is a magnetic insulator with space group Fd$\bar{3}$m in which the Gd$^{3+}$ ions occupy the sites
 of a pyrochlore lattice. The magnetic levels have $S=7/2$ and $L=0$ so the crystal electric field has an 
effect only through the mixing of states outside these $L=0$ levels. The material has a negative Curie-Weiss 
temperature $\theta_{\rm CW}=-8.6$ K and undergoes a first order transition at about $1$ K to long range order \cite{Wills} 
with a Palmer-Chalker magnetic structure \cite{PalmerChalker} and an ordered moment of about $7\mu_{B}$ \cite{Wills}. 
This magnetic structure strongly suggests that the dominant magnetic interactions are antiferromagnetic exchange
 with additional (perturbative) magnetostatic dipolar interactions. These observations as well as, 
for example, specific heat \cite{Quilliam} and electron spin resonance (ESR) measurements \cite{Sosin} have 
allowed for a determination of the exchange coupling as well as a more refined picture of the microscopics to include a 
weak XY-like single ion anisotropy. The leading contribution to the single ion Hamiltonian is
\begin{equation} H_{\rm aniso} =
 \vert\Delta\vert \sum_{\mu,a} \left(\mathbf{S}_{\mu,a}\cdot\mathbf{\hat{z}}_{a}  
 \right)^{2}
 \label{eqn:Haniso} 
\end{equation}
where $\mu$ labels the face-centred cubic space lattice sites and $a$ runs over the tetrahedral primitive basis 
of the pyrochlore lattice. The local sublattice $\mathbf{\hat{z}}_{a}$ axis is oriented in the local $[ 111]$
 direction on sublattice $a$. The anisotropy coupling $\Delta$ has been estimated to be about $0.14$ K \cite{Glazkov}. 
Further contributions to the crystal field of \gso\ are smaller and have been neglected \cite{Glazkov}.
The magnetic interaction Hamiltonian is
\begin{equation}  H_{\rm int} = \mathcal{J}_{{\rm ex}}\sum_{\langle i,j\rangle}
\mathbf{J}_{i}\cdot\mathbf{J}_{j} + \mathcal{D}r_{{\rm nn}}^{3}\sum_{{\rm i>j}}
[\mathbf{J}_{i}\cdot\mathbf{J}_{j} -
3(\mathbf{J}_{i}\cdot\mathbf{\hat{R}}_{ij})(\mathbf{J}_{j}\cdot\mathbf{\hat{R}}_{ij})]|\mathbf{R}_{ij}|^{-3}  \label{eqn:Hint} \end{equation}
where $\mathbf{R}_{ij}$ is the position vector of ion $i$ with respect to ion $j$. 
The estimate of $\mathcal{J}_{\rm ex}=0.27$ K which comes directly from $\theta_{\rm CW}$ \cite{Quilliam}
 provides an adequate fit to the ESR data in Ref.~\cite{Sosin} which measures the uniform contribution
 to the spin wave spectrum. 
The dipolar coupling $\mathcal{D}=(\mu_{0}/4\pi)(g\mu_{B})^{2}/r_{\rm nn}^{3}=0.049$ K 
where $r_{\rm nn}=\sqrt{2}a/4$ is the distance between nearest neighbor Gd$^{3+}$ ions 
and $a$ is the unit cell edge length of $10.46\hspace{1pt} \AA{}$ \cite{Quilliam}
and $g=2$ is the Gd$^{3+}$ Land\'e factor.
One can also incorporate exchange couplings beyond nearest neighbour.
 An estimate for these couplings exists \cite{delMaestro} but we neglect them
 here because they do not significantly affect our results.

The first experiment to explore the electronic spin dynamics in \gso\ was a M\"{o}ssbauer
 spectroscopy experiment which concluded that the Gd$^{3+}$ moments continue to fluctuate 
at least down to $27$ mK. The two $\mu$SR experiments of Bonville {\it et al. } \cite{Bonville}
 and Chapuis {\it et al.} \cite{Chapuis} report asymmetry data below $T_{c}$ with two clear 
oscillation signals for time $t\lesssim 0.3\mu$s corresponding to a pair of local fields of
 $207(1)$ mT and $442(1)$ mT. The initial measured asymmetry below $T_{c}$ in Ref.~\cite{Chapuis} 
is reported to be only about one quarter of that above $T_{c}$ which is perhaps due to muons at other
 stopping sites and experiencing larger internal fields giving precession signals which are outside 
the (time) resolution of the spectrometer. 
The relaxation rate $\lambda$ of the nonoscillating part of the
asymmetry shows, in both experiments, as the temperature is lowered, 
a clear and sharp peak at $T_{c}$ followed by a plateau at about $0.6 
\mu$s$^{-1}$ extending down to the lowest observed temperature of $0.02$ K.

\section{Field Distribution in Gd$_{2}$Sn$_{2}$O$_{7}$}
\label{sec:statics}

The internal fields determined from the $\mu$SR spectrum in two experiments \cite{Bonville,Chapuis} 
are very similar suggesting that the muon location is rather well-defined in \gso. Both experiments 
find a satisfactory fit to the asymmetry oscillations using two distinct fields. 
This is perhaps unexpected since, even allowing for a fixed muon location, there are three 
distinct Palmer-Chalker magnetic domains (up to time reversal).  In this section, 
we explore the magnetic fields in the vicinity of the oxygen ions, 
where the positive muon is expected to hydrogen-bond, 
assuming that (i) 
the only field that the muons experience comes from the static Gd$^{3+}$ magnetic moments and 
that (ii) the muon does not produce a local lattice distortion. Taking a cubic cell of volume 
$(2a)^{3}$ with magnetic moments in the Palmer-Chalker state \cite{PalmerChalker},
 we compute the magnetic field within spherical shells of inner radius $1 \AA{}$ 
and outer radius $1.5 \AA{}$ about each oxygen ion. 
We find locations where the three possible symmetry-related Palmer-Chalker magnetic domains give 
rise to two different fields within experimental precision which have the same magnitudes observed experimentally. 
Specifically, in units of $a$, one of these locations is $(0.4402,0.5005,0.5627)$ where the three
 Palmer-Chalker domains produce fields of $206$ mT, $207$ mT and $441$ mT. 
Since two of the fields are identical to within experimental precision, one would expect that
 the weight that field contributes to the oscillatory part of the asymmetry would be twice that 
of the other. However, the analysis of the asymmetry \cite{Chapuis} finds equal weights for the two oscillatory components. 

\section{Calculation of $\lambda$}
\label{sec:dynamics}

In this section, we compute the relaxation rate of the muon due to a two-magnon Raman process. 
This is identical to calculations of spin lattice relaxation in NMR \cite{Moriya}. 
The muon down spin ($\vert\hspace{-3pt} \downarrow\rangle$) to up spin ($\vert\hspace{-3pt}\uparrow\rangle$) transition
 rate is given by the Fermi Golden rule
 \begin{equation} W = \frac{2\pi}{\hbar}
\sum_{a,b,{\rm i},{\rm f}}\langle \left| \langle {\rm f};\uparrow | H_{\rm \mu -Gd} |{\rm i};\downarrow\rangle  
 \right|^{2}  \rangle_{\rm th} \delta(\omega_{\rm f} -\omega_{\rm i} - \gamma_{\mu}H)  
 \label{eqn:FGR} 
 \end{equation}
 where a sum is taken over all initial (i) and final (f) magnon states and $\gamma_{\mu}H$
 is the muon Zeeman splitting in the local field $H$. $\langle\ldots \rangle_{\rm th}$ denotes a thermal average.

The reference (unperturbed) Hamiltonian of \gso\ is the sum of Eqns.~(\ref{eqn:Haniso}) and (\ref{eqn:Hint})
 and the interaction between the magnetically ordered moments and the muon spin $\mathbf{I}$ 
is an interaction between dipole moments given by
\[  \fl H_{\rm \mu-Gd} = \left(\frac{\mu_{0}}{4\pi}\right)g_{\rm J}\mu_{B}
 \gamma_{\mu}\hbar\sum_{\mu,a}\frac{1}{|\mathbf{R}^{\mu}_{a}|^{3}}\left\{\left( \mathbf{I}\cdot
 \delta \mathbf{S}_{a}(\mathbf{R}^{\mu})\right)
 - 3\left( \mathbf{I}\cdot\mathbf{\hat{R}}_{a}^{\mu} \right)
\left(   \delta\mathbf{S}_{a}(\mathbf{R}^{\mu})\cdot\mathbf{\hat{R}}_{a}^{\mu} \right)    \right\}.   
     \]
We have included only the deviation of the moments from their ordered (Palmer-Chalker)
 directions $\delta \mathbf{S}$ because only the fluctuation part can lead to dynamical
 relaxation of an individual muon. The static part of the interaction Hamiltonian, which
 is not shown, causes the muon to precess in the local field and a static distribution of such
internal fields produces the nondynamical relaxation described in the Introduction. 
To proceed, we have to consider the possible processes that might give rise to relaxation.
 We work in the linear spin wave approximation using Holstein-Primakoff bosons \cite{Sosin,delMaestro,delMaestro2}. 
The simplest process is one where a single magnon is created or annihilated, accompanied by a muon spin flip.
 Such a process appears to lowest order from the transverse spin components, for these are linear 
in the magnon creation and annihilation operators but it is forbidden by energy 
conservation because the spin wave gap is much larger than the muon Zeeman splitting per unit field.
 To illustrate this point, we compute the Zeeman splitting of a muon 
to be $\hbar\gamma_{\mu}=0.0065$ KT$^{-1}$. Since the local field is of the order of $10^{2}$ mT \cite{Bonville,Chapuis},
the Zeeman energy is about $6.5\times 10^{-4}$ K,
 while the spin wave gap is about $0.5$ K in Gd$_2$Sn$_2$O$_7$ \cite{Sosin,delMaestro,delMaestro2}, 
so a single magnon typically has an energy that exceeds the muon spin flip energy by three orders of magnitude.
 For this reason, we neglect the transverse Gd$^{3+}$ spin fluctuations and drop the Zeeman energy
 from the energy conserving delta function in Eqn.~(\ref{eqn:FGR}). 
With only longitudinal fluctuations contributing to the relaxation, let
\begin{equation}  f^{a}(\mathbf{R}^{\mu}) \equiv \left(\frac{\mu_{0}}{4\pi}\right)g_{\rm J}\mu_{B} 
\gamma_{\mu}\hbar\frac{1}{|\mathbf{R}^{\mu}_{a}|^{3}}\left\{\left( \mathbf{I}\cdot \mathbf{z}_{a}\right) -
 3\left( \mathbf{I}\cdot\mathbf{\hat{R}}_{a}^{\mu} \right)\left(   \mathbf{z}_{a}\cdot\mathbf{\hat{R}}_{a}^{\mu} \right)    
\right\}  \label{eqn:f}   \end{equation}
where unit vectors $\mathbf{\hat{z}}_{a}$ are along the local quantization axis of the Gd$^{3+}$ spins 
and depend only on $a$ because the ordering wavevector is $\mathbf{q}=0$ in the Palmer-Chalker state \cite{PalmerChalker}. 
The interaction can now be written as
\begin{equation} H_{\rm \mu-Gd} = -\sum_{\mu,a} \hat{n}_{a}(\mathbf{R}^{\mu})f^{a}(\mathbf{R}^{\mu})  
\label{eqn:HmuGd} 
\end{equation}
with $\hat{n}_{a}=c_{a}^{\dagger}(\mathbf{R}^{\mu})c_{a}(\mathbf{R}^{\mu})$ for $S^{z}_{a}=S-\delta S^{z}_{a} = S-c^{\dagger}_{a}c_{a}$ 
in terms of Holstein-Primakoff boson operators. The lowest order energy conserving contribution to the relaxation rate is a 
two magnon Raman process whereby the muon spin is flipped via the scattering of a magnon. The matrix element for this process
 is
\begin{equation}    \mathcal{M}_{aa'bb'}(\mathbf{k},\mathbf{k}') =
 \langle n_{a'}(\mathbf{k})+1,n_{b'}(\mathbf{k}')-1;\uparrow | H_{\rm \mu-Gd}  
|n_{a}(\mathbf{k}),n_{b}(\mathbf{k}');\downarrow\rangle  
\label{eqn:ME} \end{equation}
where $n_{a}$ is the boson occupation number for  magnon branch $a$, with 
$a$ running over the four magnetic sublattices, $a=1,\ldots,4$, 
of the Palmer-Chalker state \cite{PalmerChalker}.

\begin{figure}
\begin{center}
\subfigure{
\includegraphics[scale = 0.25]{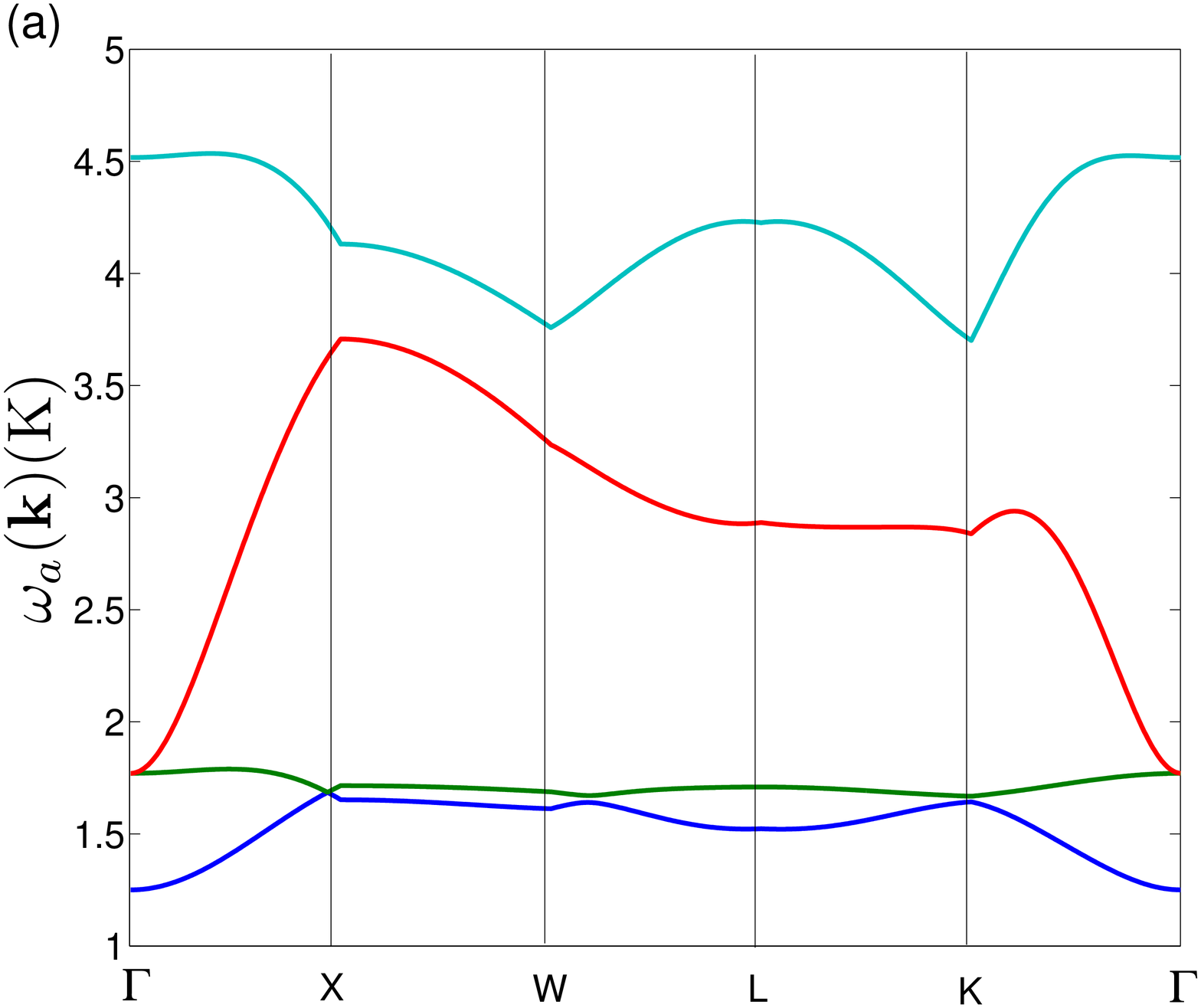}
\label{fig:SWa}
}
\subfigure{
\includegraphics[scale=0.25]{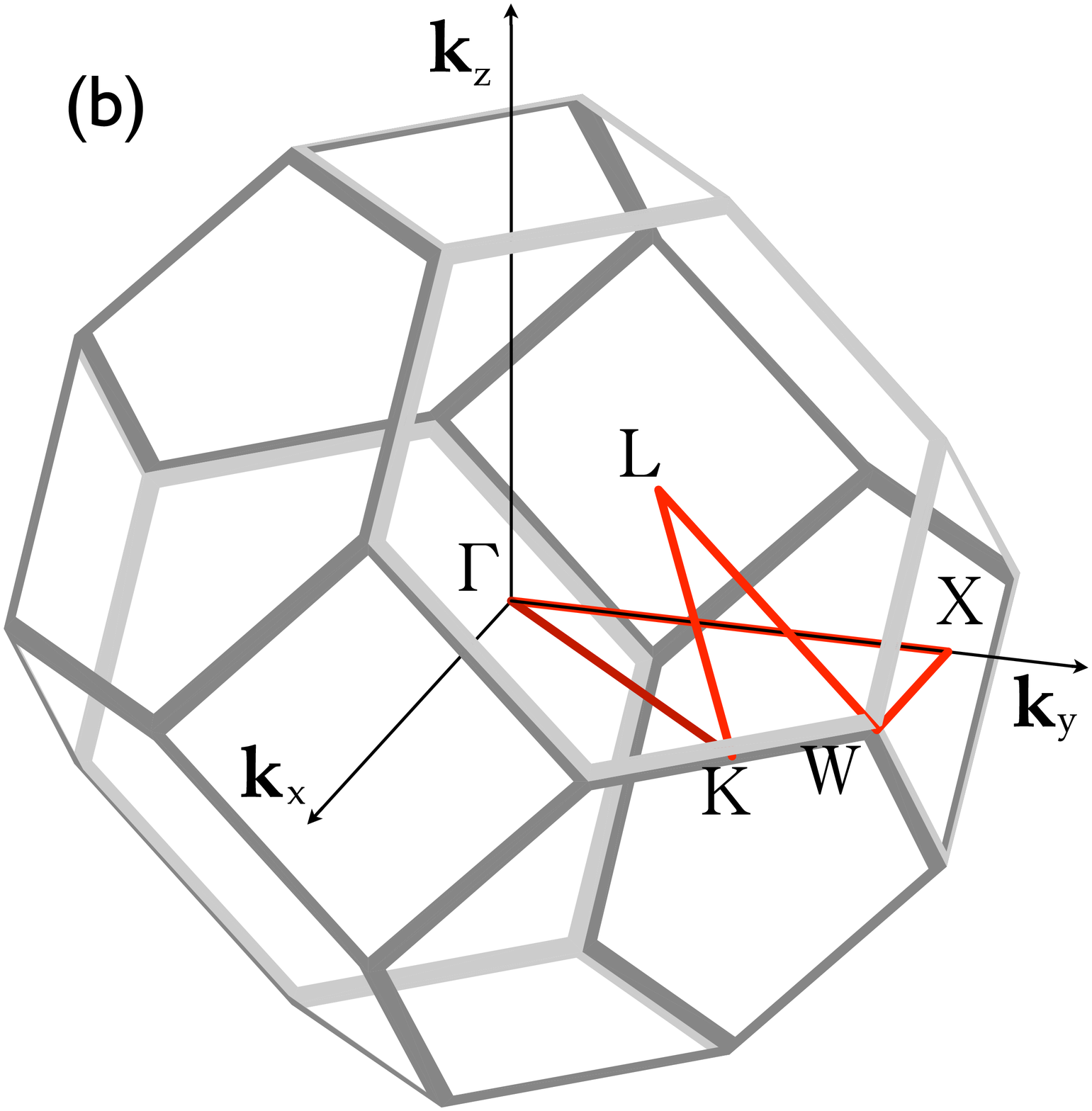}
\label{fig:SWb}
}
\caption{Panel \ref{fig:SWa} shows the spin waves frequencies, $\omega_a({\mathbf k})$
for \gso\ along paths in the Brillouin zone of 
the BCC reciprocal lattice of the FCC space lattice shown in Panel~\ref{fig:SWb}. The magnon gap never reaches its minimum of about $0.7$ K along this path.}
\label{fig:spinwaves}
\end{center}
\end{figure}

The calculation of the spin wave spectrum for this material has been discussed elsewhere \cite{Quilliam,Sosin,delMaestro,delMaestro2} 
and we refer to the literature for further details. We replace the spins in the Hamiltonian $H\equiv H_{\rm aniso}+H_{\rm int}$  
with Holstein-Primakoff bosons and expand in powers of $1/S$ up to and including powers of $S^{0}$ which gives a harmonic Hamiltonian. 
One finds that the terms linear in the boson creation and annihilation operators, $c^{\dagger}$ and $c$,
 vanish in the Palmer-Chalker spin configuration, as required of a semi-classical ground state of the Hamiltonian $H$.
 The $c^{\dagger}$,$c$ operators are transformed via a canonical transformation $Q^{ab}(\mathbf{k})$, 
an $8\times 8$ matrix, into spin wave operators $a^{\dagger}$ and $a$.
 The transformation,
\begin{eqnarray*} 
c_{a}(\mathbf{k}) 		
 &=& \sum_{b=1}^{4} Q^{ab}_{\mathbf{k}}a_{b}(\mathbf{k}) + Q^{ab+4}_{\mathbf{k}}a^{\dagger }_{b}(-\mathbf{k}) \\
c_{a}^{\dagger}(-\mathbf{k}) 
 &=& \sum_{b=1}^{4} Q^{a+4b}_{\mathbf{k}}a_{b}(\mathbf{k}) + Q^{a+4b+4}_{\mathbf{k}}a^{\dagger }_{b}(-\mathbf{k}) ,
\end{eqnarray*}
where index $a$ runs over the sublattices,  is such that the spin wave operators also satisfy Bose commutation relations
 and transform the spin Hamiltonian into the form
\[  H =  H^{(0)} + \frac{\hbar}{2}\sum_{\mathbf{k}}\sum_{a} \omega_{a}(\mathbf{k}) + 
\sum_{\mathbf{k}}\sum_{a}\hbar\omega_{a}(\mathbf{k}) \left( a_{a}^{\dagger}(\mathbf{k})a_{a}(\mathbf{k}) \right)  , \]
where the spin wave frequencies are $\omega_{a}(\mathbf{k})$. The spectrum along line segments between points 
of high symmetry in the Brillouin zone is shown in Fig.~\ref{fig:spinwaves} where the canonical transformation
 has been computed numerically \cite{delMaestro,delMaestro2}. 

\begin{figure}
\begin{center}
\subfigure{
\includegraphics[scale = 0.22]{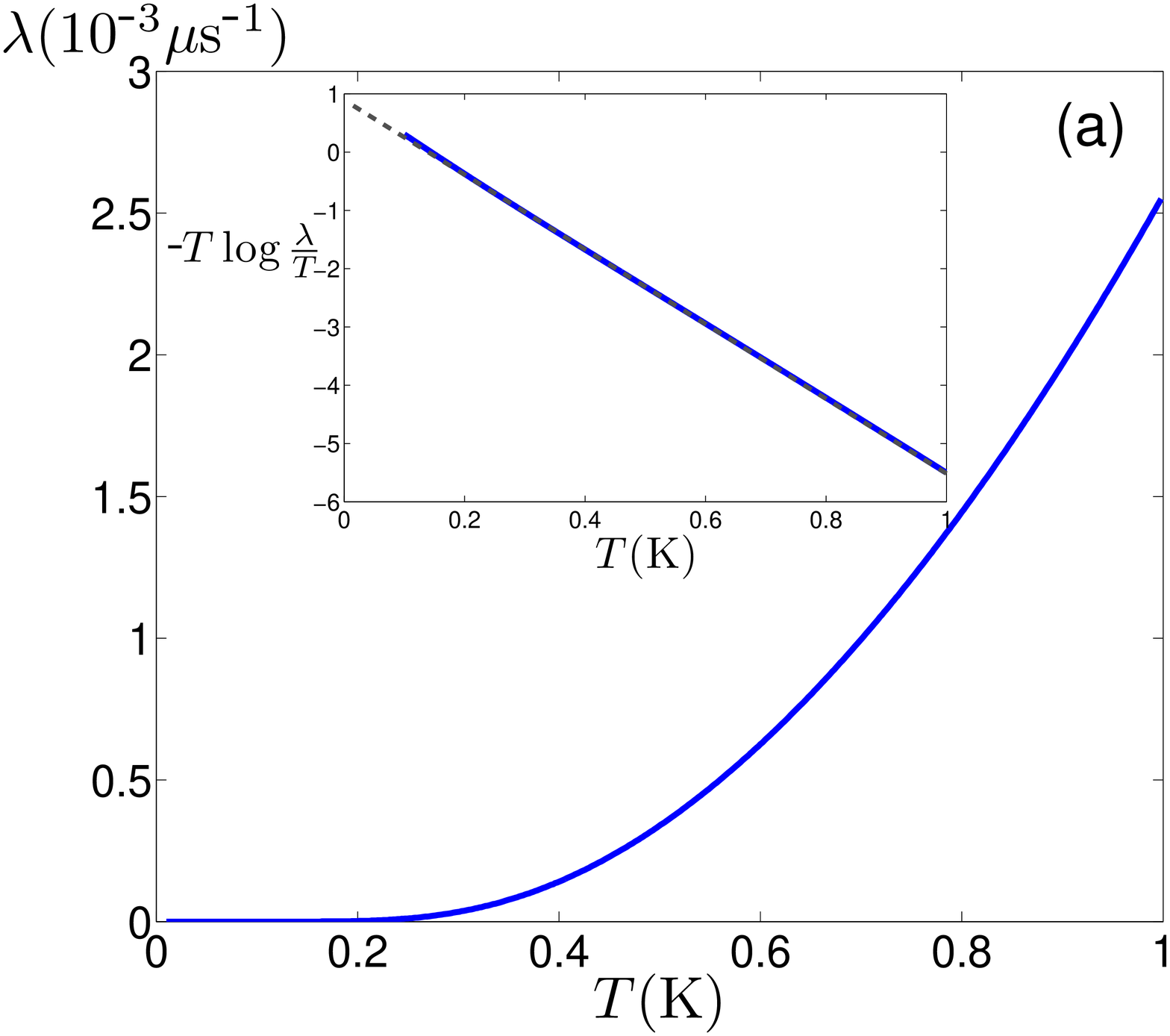}
\label{fig:lambdaT}
}
\subfigure{
\includegraphics[scale=0.22]{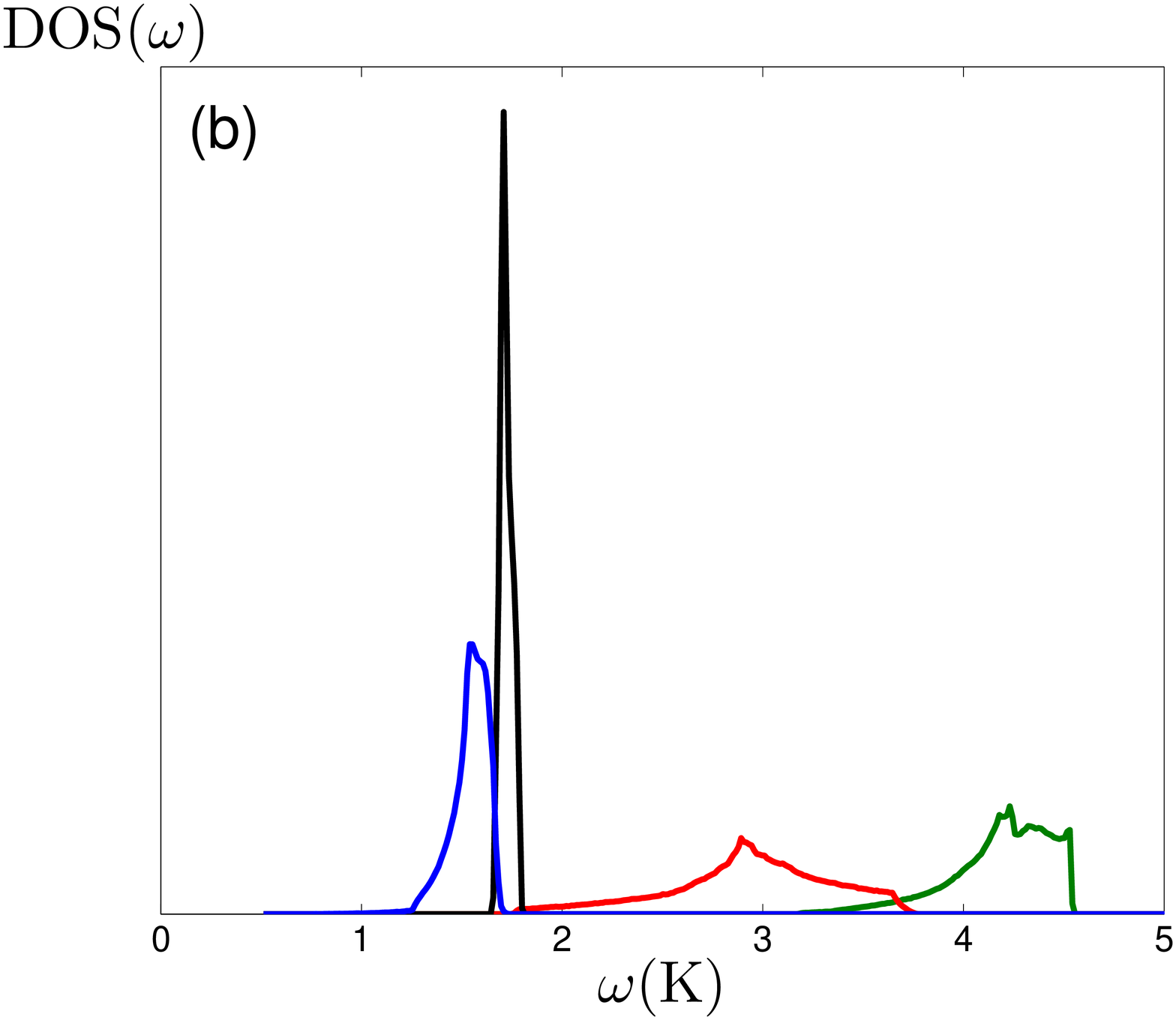}
\label{fig:omegaDOS}
}
\caption{Panel (a) is the two magnon Raman muon $\lambda(T)$ for \gso\ and Panel (b) is a plot of
 the densities of states (DOS) for each of the four branches of the spin wave spectrum. The lowest branch in the density of states has a low energy tail that extends down to the spectral gap of about $0.7$ K.}
\label{fig:rate}
\end{center}
\end{figure}

We can now compute the matrix element  
$\mathcal{M}_{aa'bb'}(\mathbf{k},\mathbf{k}')$ of Eqn.~(\ref{eqn:ME}) using
$a^{\dagger}(\mathbf{k}) | n(\mathbf{k})\rangle = \sqrt{n(\mathbf{k})+1} |n(\mathbf{k})+1\rangle$ 
and 
$a(\mathbf{k}) | n(\mathbf{k})\rangle = \sqrt{n(\mathbf{k})} |n(\mathbf{k})-1\rangle$. 
We find
\begin{eqnarray*}
\eqalign \fl \mathcal{M}_{aa'bb'}(\mathbf{k},\mathbf{k}') 
& = \delta_{aa'}\delta_{bb'} \sum_{\mu,c} 
 \left[ ( Q^{\star}_{\mathbf{k}})^{ca}Q^{cb}_{\mathbf{k}'} 
+ (Q^{\star}_{-\mathbf{k}'})^{cb+4}Q^{ca+4}_{-\mathbf{k}} \right ] \\ 
& \times \sqrt{n_{a}(\mathbf{k})+1}\sqrt{n_{b}(\mathbf{k}')}\exp(i(\mathbf{k}-\mathbf{k}')\cdot\mathbf{R}_{c}^{\mu}) 
\langle \uparrow \vert f^{c}(\mathbf{R}^{\mu})\vert \downarrow\rangle
\end{eqnarray*}
where $Q^{\star ab}$ is the complex conjugate of $Q^{ab}$.
Computing the squared modulus of this matrix element, we make the approximation that the phase factor 
$\exp(i(\mathbf{k}-\mathbf{k}')\cdot(\mathbf{R}_{c}^{\mu}-\mathbf{R}_{d}^{\nu}))$ averages
 to zero unless $\mathbf{R}^{\mu}_{c}-\mathbf{R}^{\nu}_{d}=0$ so that
\[ \fl \vert\mathcal{M}\vert^{2} \approx \frac{1}{N^{2}} \sum_{c}  n_{b}(\mathbf{k}') (n_{a}(\mathbf{k})+1)F_{c}
 \left(Q^{\star ca}_{\mathbf{k}}Q^{cb}_{\mathbf{k}'}
 + Q^{\star cb+4}_{-\mathbf{k}'}Q^{ca+4}_{-\mathbf{k}} \right)\left( Q^{ca}_{\mathbf{k}}Q^{\star cb}_{\mathbf{k}'} 
+ Q^{cb+4}_{-\mathbf{k}'}Q^{\star ca+4}_{-\mathbf{k}}  \right) , \]
where $F^{c}\equiv \sum_{\mu} \vert \langle \uparrow\vert f^{c}(\mathbf{R}^{\mu})\vert \downarrow \rangle\vert^{2}$.
The transition rate $W$ in Eqn.~(\ref{eqn:FGR})  then becomes
\begin{eqnarray*} \fl W = \frac{2\pi}{\hbar} \frac{1}{\Omega^{2}_{\rm BZ}} 
\int_{\rm BZ}d\mathbf{k}\int_{\rm BZ} d\mathbf{k}' 
\sum_{a,b,c} n_{b}(\mathbf{k}')(n_{a}(\mathbf{k})+1)
 \delta(\omega_{a}(\mathbf{k})-\omega_{b}(\mathbf{k}')) F_{c} 
\left\{ (Q_{\mathbf{k}}^{ca}Q_{\mathbf{k}}^{\star ca})(Q_{\mathbf{k}'}^{cb}Q_{\mathbf{k}'}^{\star cb}) \right. \\ \left. 
\fl +(Q_{-\mathbf{k}}^{ca+4}Q_{-\mathbf{k}}^{\star ca+4})(Q_{-\mathbf{k}'}^{cb+4}Q_{-\mathbf{k}'}^{\star cb+4}) +(Q_{\mathbf{k}}^{\star ca}Q_{-\mathbf{k}}^{\star ca+4})(Q_{\mathbf{k}'}^{cb}Q_{-\mathbf{k}'}^{cb+4})+(Q_{-\mathbf{k}}^{ca+4}Q_{\mathbf{k}}^{ca})(Q_{-\mathbf{k}'}^{\star cb+4}Q_{\mathbf{k}'}^{\star cb})  
\right\}  ,
\end{eqnarray*}
where the thermal average at inverse temperature $\beta$ of the boson occupation numbers
 is the distribution $n_{a}(\mathbf{k})=(\exp(\beta\omega_{a}(\mathbf{k}))+1)^{-1}$.
The integrals can be regrouped into a single energy integral by introducing densities of states,
\begin{eqnarray}
N_{1}^{cb}(\omega) & 
 \equiv & \frac{1}{\mathcal{N}}\int_{\omega} 
d\mathbf{k} Q_{\mathbf{k}}^{cb}Q_{\mathbf{k}}^{\star cb} 
\hspace{15pt} 
N_{2}^{cb}(\omega)  
 \equiv  \frac{1}{\mathcal{N}}\int_{\omega} 
d\mathbf{k} Q_{-\mathbf{k}}^{cb+4}Q_{-\mathbf{k}}^{\star cb+4} \nonumber \\
N_{3}^{cb}(\omega) &  
\equiv  & \frac{1}{\mathcal{N}}\int_{\omega} 
d\mathbf{k} Q_{\mathbf{k}}^{cb}Q_{-\mathbf{k}}^{cb+4} 
\hspace{15pt} N_{4}^{cb}(\omega)\equiv \frac{1}{\mathcal{N}}\int_{\omega} d\mathbf{k} 
Q_{-\mathbf{k}}^{cb+4}Q_{\mathbf{k}}^{cb} = N_{3}^{cn}(\omega) ,
\label{eqn:DOS}
\end{eqnarray}
where the $\omega$ indicates that the integrals are evaluated for fixed energy and the indices $b$ and $c$ run over 
sublattices $1$ to $4$.
The normalization factor $\mathcal{N}$ is chosen to ensure that $(1/\mathcal{N})\int_{\omega} d\mathbf{k}=1$. 
Finally, the transition rate $W$ is given by
\begin{equation}  W = \left(\frac{2\pi}{\hbar}\right)
 \sum_{m=1}^{4}\sum_{a,b,c} \int d\omega \hspace{1pt} n(\omega)(n(\omega)+1) F^{c} N_{m}^{ca\star}N_{m}^{cb}  
 \label{eqn:rate} \end{equation}
It remains for us to evaluate $F_{c}$ from Eqn.~(\ref{eqn:f}). 
Because the \gso\ sample in Refs.~\cite{Bonville,Chapuis} is a powder, 
we must average over the crystal orientations with respect to the initial muon polarization axis.
 For this purpose, it is convenient to define three separate coordinate systems: (i)
 the muon polarization system $\mbox{\boldmath$\hat{\mu}$}$, (ii) the spin polarization system on 
each sublattice $\mathbf{\hat{S}}_{a}$ and (iii) a system about the axis $\mathbf{\hat{R}}^{\mu}_{a}$.
 We introduce rotation matrices $U_{a}^{\alpha\beta}$ and $V^{\alpha\beta}_{a\mu}$ 
such that 
$(\mathbf{\hat{S}}_{a})^{\alpha} \equiv U_{a}^{\alpha\beta}
\mbox{\boldmath$\hat{\mu}$}_{\beta}$
and 
$(\mathbf{\hat{S}}_{a})^{\alpha} \equiv V_{a\mu}^{\alpha\beta}(\mathbf{\hat{R}}^{\mu}_{a})_{\beta}$. 
These allow to write $f^{c}(\mathbf{R}^{\mu})$ as
\[ \fl f^{c}(\mathbf{R}^{\mu}) =  \frac{1}{2} 
\left(\frac{\mu_{0}}{4\pi}\right)g_{\rm J}\mu_{B}
 \gamma_{\mu}\hbar \frac{1}{\vert\mathbf{R}^{\mu}_{c}\vert^{3}} \left\{ \left(U_{c}^{xz}-iU_{c}^{yz}\right) 
- 3V_{c\mu}^{zz} \sum_{\beta} V^{\beta z}_{c\mu}\left( U_{c}^{\beta x} - iU_{c}^{\beta y} \right) \right\}  . \]
Performing the angular average, we obtain
\[  \fl F^{c} = \frac{1}{2} \left(\frac{\mu_{0}}{4\pi}\right)^{2} (g_{\rm J}\mu_{B} \gamma_{\mu}\hbar)^{2} 
\sum_{\mu}\frac{1}{\vert\mathbf{R}^{\mu}_{c}\vert^{6}} \left\{ \frac{1}{3} + 3\left(V_{c\mu}^{zz}\right)^{2}
\left( (V_{c\mu}^{xz})^{2} + (V_{c\mu}^{yz})^{2} + (V_{c\mu}^{zz})^{2} \right) \right\} . \]
We evaluate Eq.~(\ref{eqn:rate}) averaged over the three possible Palmer-Chalker magnetic domains 
and over all crystal orientations. The densities of states $N_m^{cb}(\omega)$ in Eqn.~(\ref{eqn:DOS})
were computed by Monte Carlo integration 
by introducing energy bins over the range of magnon energies. For points $\mathbf{k}$
 distributed uniformly in the first Brillouin zone,  $\omega_{a}(\mathbf{k})$ are determined whereupon 
the integrand in the density of states is added into the corresponding energy bins. 
The muon location was taken to be the position given in Section~\ref{sec:statics}.
Finally, the muon polarization relaxation rate $\lambda=2W$, 
where the factor $2$ arises from the fact that both spin-up
 and spin-down populations contribute from the relaxation rate \cite{Moriya}, is shown in 
Fig.~\ref{fig:rate} plotted as a function of temperature up to the transition temperature. 
 Compared to the experimental plateau at $0.6\mu{\rm s}^{-1}$, the two magnon Raman relaxation rate
 is very small - about one-hundredth of the experimental relaxation rate in the vicinity of $T_{c}$.
 In addition, the relaxation rate is strongly temperature dependent. 
A calculation of the relaxation rate for a gapped cubic lattice antiferromagnet \cite{Moriya} 
gives a trend $\lambda \propto \lambda_0  T\exp(-\Delta/T)$ which is illustrated for our problem in the inset
 to Panel (a) of Fig.~\ref{fig:rate}. 
The $T\rightarrow 0$ limit of $\lambda(T)$ in the inset is consistent with the spectral gap $\Delta\sim 0.7$ K.
 
\section{Discussion and Conclusions}
\label{sec:disc}

We have computed the relaxation rate of the muon spin polarization due to the scattering of a pair 
of magnons in the ordered phase of \gso. As far as we are aware, this is
 the first such detailed microscopic calculation of $\mu$SR $\lambda(T)$ 
for a frustrated magnet exhibiting PSD. We confirm explicitly that the rate is strongly temperature dependent,
 in line with the observation that the magnon density should fall off exponentially at temperatures less 
than the spectral gap $\Delta\approx 0.7$ K. The two-magnon relaxation rate is a strongly 
temperature dependent background to the experimentally measured $\lambda$. However, in 
spite of the strong temperature dependence, we find that $\lambda$ just below $T_{c}$ is smaller than the experimental rate by more than two orders of magnitude. Since
the variation of the computed rate with
 temperature is smaller than the experimental resolution, the two-magnon (Raman) process in 
\gso\  is completely hidden by whatever mechanism is responsible for the 
$\lambda\sim 0.6$ $\mu$s$^{-1}$ relaxation rate plateau.

The relaxation of the muon polarization of the order of $0.6$ $\mu$s$^{-1}$ is therefore entirely inexplicable 
over the {\it whole} temperature range below $T_c \approx 1$ K within the 
usual context of spin lattice relaxation. This result puts aside the possibility that the cause of this phenomenon comes 
from unusual features of the spin wave spectra of geometrically frustrated magnets \cite{Yaouanc,Bonville2}. 
Higher order magnon processes are also suppressed at temperatures below the spin wave gap \cite{Beeman} 
so neither the plateau nor the $\lambda\sim 0.6$ $\mu$s$^{-1}$ rate can be explained by considering the 
scattering of conventional (gapped) magnons. It is interesting, in light of this result, 
to recall details of some suggestive experimental findings.

Firstly, PSD is typically regarded as a distinctive feature of magnetic materials on frustrated lattices 
as opposed to conventional low temperature magnets. However, beyond this feature, there are not many similarities between 
the materials exhibiting anomalous low temperature relaxation. Notably, one finds PSD in materials without 
long range order (e.g. 
 Tb$_{2}$Ti$_{2}$O$_{7}$ \cite{Gardner,Keren}
and Yb$_{2}$Ti$_{2}$O$_{7}$ \cite{Hodges}).
 Indeed, PSD uncovered by $\mu$SR is often taken to be a good evidence from a local probe of spin liquid behaviour.
 It is not clear that the mechanism underlying the spin dynamics is the same across the large 
spectrum of materials exhibiting PSD. However, if the mechanism is common to all these materials, 
then its manifestation in collective paramagnetic or spin liquid
phases may constitute further evidence that magnons are 
irrelevant to low temperature $\lambda$ plateaux. 
An interesting finding is that the $\lambda$ plateau is 
little affected by doping Tb$_{2}$Ti$_{2}$O$_{7}$ 
with nonmagnetic impurities  \cite{Keren}, suggesting that the plateau, at least in this compound, 
is not due to conventional collective excitations. 
To emphasize, the observation of PSD in (gapless) spin liquid systems is not our
immediate concern. Rather, it is the occurence of PSD in pyrochlores such
as  
Tb$_{2}$Sn$_{2}$O$_{7}$ \cite{deReotier,Bert_TSO,Giblin,Bonville,ChapuisTSO,RuleTSO1,MirebeauTSO,RuleTSO2},
Er$_{2}$Ti$_{2}$O$_{7}$ \cite{Lago},  
Gd$_{2}$Ti$_{2}$O$_{7}$ \cite{Dunsiger1,Yaouanc}
Gd$_{2}$Sn$_{2}$O$_{7}$ \cite{Bonville,Chapuis}
as well
as spinels \cite{Kalvius}
 such as FeAl$_2$O$_4$ and CoAl$_2$O$_4$ that display long range order as exposed by
neutron scattering experiments that we believe is particularly paradoxical.

One might wonder whether the muon itself might be 
contributing to this phenomenon. In this context
it is worth commenting on 
Tb$_{2}$Sn$_{2}$O$_{7}$.
In this material, spin dynamics below $T_{c}$ has been observed in 
$\mu$SR \cite{deReotier,Bert_TSO,Bonville}
 as well as with a neutron probe 
\cite{ChapuisTSO,RuleTSO1,MirebeauTSO,RuleTSO2}, albeit on much shorter timescales
from those in $\mu$SR experiments and apparently with a significant temperature dependence \cite{MirebeauTSO}.
In conclusion, the microscopic origin of the ubiquitous phenomenon of PSD in highly frustrated
magnetic materials remains an important and unsolved problem.

\ack

We thank F. Mila, D. Poilblanc and
 T. Yavors'kii for useful discussions.
 This research was funded by the NSERC of Canada and the Canada Research Chair program (M. G., Tier I).


\Bibliography{99}
\bibitem{MuonReview} Blundell S J {et al.} 1999 Spin polarized muons in condensed matter systems {\it Contemporary Physics }{\bf 40} 175 
\bibitem{MuonReview2} Carretta P and Keren A 2009 NMR and muSR in Highly Frustrated Magnets {\it Preprint} arXiv:0905.4414

\bibitem{Bert} Bert F, Mendels P, Bono D, Olariu A, Ladieu F, Trombe J-C, Duc F, Baines C, Amato A, Hillier A 2006 Dynamics in pure and substituted volborthite kagome-like compounds {\it Physica B} {\rm 374} 134
\bibitem{Fudamoto} Fudamoto Y {\it et al.} 2003 Muon spin relaxation and susceptibility studies of the pure and diluted spin $\frac{1}{2}$ Kagome-like lattice system $($Cu$_{x}$Zn$_{1-x})_{3}$V$_{2}$O$_{7}$(OH$_{2}$)$2$H$_{2}$O {\it Phys. Rev. Lett.} {\rm 91} 207603

\bibitem{Keren3} Keren A {\it et al.} 1996 Muon spin rotation measurements in the kagome lattice systems: Cr-jarosite and Fe-jarosite {\it Phys. Rev. B} {\bf 53} 6451

\bibitem{Zorko} Zorko {\it et al.} 2008 Easy-axis kagome antiferromagnet: local-probe study of Nd$_{3}$Ga$_{5}$SiO$_{14}$ {\it Phys. Rev. Lett.} {\bf 100} 147201

\bibitem{Uemura} Uemura {\it et al.} 1994 Spin fluctuations in frustrated kagome lattice system SrCr$_{8}$Ga$_{4}$O$_{19}$ studied by muon spin relaxation {\it Phys. Rev. Lett.} {\bf 73} 3306
\bibitem{Keren2} Keren A {\it et al.} 2000 Magnetic dilution in the geometrically frustrated SrCr$_{9p}$Ga$_{12-9p}$O$_{19}$ and the role of local dynamics: a muon spin relaxation study {\it Phys. Rev. Lett.} {\bf 84} 3450 

\bibitem{Lago} Lago J {\it et al.} 2005 Magnetic ordering and dynamics in the XY pyrochlore antiferromagnet: a muon spin relaxation study of Er$_{2}$Ti$_{2}$O$_{7}$ and Er$_{2}$Sn$_{2}$O$_{7}$ {\it J. Phys.:Condens. Matter} {\bf 17} 979 

\bibitem{Dunsiger1} Dunsiger S R {\it et al.} 2006
Magnetic field dependence of muon spin relaxation in geometrically frustrated Gd$_2$Ti$_2$O$_7$
Phys. Rev. B {\bf 73}, 172418.

\bibitem{Yaouanc} Yaouanc A {\it et al.} 2005 Magnetic density of states at low energy in geometrically frustrated systems {\it Phys. Rev. Lett.} {\bf 95} 047203



\bibitem{deReotier} Dalmas de R\'{e}otier P {\it et al.} 2006 
Spin Dynamics and Magnetic Order in Magnetically Frustrated Tb$_{2}$Sn$_{2}$O$_{7}$ {\it Phys. Rev. Lett.} {\bf 96} 127202

\bibitem{Bert_TSO} Bert F {\it et al.} (2006)
Direct Evidence for a Dynamical Ground State in the Highly Frustrated Tb$_2$Sn$_2$O$_7$  Pyrochlore
Phys. Rev. Lett. {\bf 97}, 117203 (2006).

\bibitem{Giblin} Giblin S R {\it et al.} 2008 Static Magnetic Order in Tb$_{2}$Sn$_{2}$O$_{7}$ Revealed by Muon Spin Relaxation with Exterior Muon Implantation  {\it Phys. Rev. Lett.} {\bf 101} 237201

\bibitem{Bonville} Bonville {\it et al} 2004 Transitions and spin dynamics at very low temperature in the pyrochlores Yb2Ti2O7 and Gd$_{2}$Sn$_{2}$O$_{7}$ {\it Hyperfine Interaction} {\bf 156/157} 103
\bibitem{Chapuis} Chapuis {\it et al.} 2009 Probing the ground state of Gd$_{2}$Sn$_{2}$O$_{7}$ through $\mu$SR measurements {\it Physica B} {\bf 404} 686

\bibitem{Kalvius} Kalvius G M {\it et al.} 2010 Frustration driven magnetic states of A-site spinels probed by $\mu$SR {\it Eur. Phys. J. B} {\bf 77} 87

\bibitem{Gardner} Gardner J S {et al.} 1999  Cooperative Paramagnetism in the Geometrically Frustrated Pyrochlore Antiferromagnet Tb$_{2}$Ti$_{2}$O$_{7}$ {\it Phys. Rev. Lett.} {\bf 82} 1012

\bibitem{Keren} Keren A {\it et al.} 2004 Dynamic Properties of a Diluted Pyrochlore Cooperative Paramagnet $($Tb$p$Y$_{1-p})_{2}$Ti$_{2}$O$_{7}$ {\it Phys. Rev. Lett.} {\bf 92} 107204

\bibitem{Hodges} Hodges J A {\it et al.} 2002 First-Order Transition in the Spin Dynamics of Geometrically Frustrated Yb$_{2}$Ti$_{2}$O$_{7}$ {\it Phys. Rev. Lett.} {\bf 88} 077204

\bibitem{Dunsiger2} Dunsiger S R {\it et al.} 1996 Muon spin relaxation investigation of the spin dynamics of geometrically frustrated antiferromagnets Y$_{2}$Mo$_{2}$O$_{7}$ and Tb$_{2}$Mo$_{2}$O$_{7}$ {\it Phys. Rev. B} {\rm 54} 9019 

\bibitem{Dunsiger} Dunsiger S R {\it et al.} 2000 Low Temperature Spin Dynamics of the Geometrically Frustrated Antiferromagnetic Garnet Gd$_{3}$Ga$_{5}$O$_{12}$ Phys. Rev. Lett., {\rm 85} 3504

\bibitem{Marshall} Marshall I M {\it et al.}  2002 A muon-spin relaxation ($\mu$SR) study of the geometrically frustrated magnets GdGaO and ZnCrO {\it J. Phys.:Condens. Matter} {\bf 14} L157 

\bibitem{Harris} Harris M J {\it et al.} 1998 Magnetic structures of highly frustrated pyrochlores {\it J. Mag. Magn. Matr.} {\bf 177} 757

\bibitem{LagoSpinIce} Lago J {\it et al.} 2007 $\mu$SR investigation of spin dynamics in the spin-ice material Dy$_{2}$Ti$_{2}$O$_{7}$ {\it J. Phys.:Condens. Matter.} {\bf 19 } 326210



\bibitem{Quilliam} Quilliam J A {\it et al.} 2007 Evidence for Gapped Spin-Wave Excitations in the Frustrated Gd$_ {2}$Sn$_ {2}$O$_ {7}$ Pyrochlore Antiferromagnet from Low-Temperature Specific Heat Measurements {\it Phys. Rev. Lett.} {\bf 99} 097201

\bibitem{Stewart} Stewart J R {\it et al.} 2008 Collective dynamics in the Heisenberg pyrochlore antiferromagnet Gd$_{2}$Sn$_{2}$O$_{7}$ {\it Phys. Rev. B} {\bf 78} 132410

\bibitem{Sosin} Sosin S S {\it et al.} 2009 Magnetic excitations in the geometrically frustrated pyrochlore antiferromagnet Gd$_{2}$Sn$_{2}$O$_{7}$ studied by electron spin resonance {\it Phys. Rev. B} {\bf 79} 014419 

\bibitem{Bertin} Bertin {\it et al.} 2002 Effective hyperfine temperature in frustrated Gd$_{2}$Sn$_{2}$O$_{7}$: two level model and $^{155}$Gd Mossbauer measurements {\it Eur. Phys. Journal B} {\bf 27} 347


\bibitem{Wills} Wills A S {\it et al.} 2006 Magnetic ordering in Gd$_{2}$Sn$_{2}$O$_{7}$: the archetypal Heisenberg pyrochlore antiferromagnet {\it J. Phys.:Condens. Matter} {\bf 18} L37

\bibitem{PalmerChalker} Palmer S E and Chalker J T 2000 Order induced by dipolar interactions in a geometrically frustrated antiferromagnet {\it Phys. Rev. B} {\bf 62} 488

\bibitem{Glazkov} Glazkov V N {\it et al.} 2006 Observation of a transverse magnetization in the ordered phases of the pyrochlore magnet Gd$_{2}$Ti$_{2}$O$_{7}$ {\it J. Phys.:Condens. Matter} {\bf 18} 2285



\bibitem{delMaestro} del Maestro A G and Gingras M J P 2007 Low-temperature specific heat and possible gap to magnetic excita
tions in the Heisenberg pyrochlore antiferromagnet Gd$_{2}$Sn$_{2}$O$_{7}$ {\it Phys. Rev. B} {\bf 76} 064418


\bibitem{Moriya} Moriya T 1956 Nuclear spin relaxation in antiferromagnets {\it Prog. Theor. Phys.} {\bf 16} 23


\bibitem{delMaestro2} del Maestro A G and Gingras M J P 2004 Quantum spin fluctuations in the dipolar Heisenberg-like rare earth pyrochlores {\it J. Phys.:Condens. Matter} {\bf 16} 3339

\bibitem{Bonville2} Bonville P {\it et al.} 2004 Transitions and Spin Dynamics at Very Low Temperature in the Pyrochlores Yb$_{2}$Ti$_{2}$O$_{7}$ and Gd$_{2}$Sn$_{2}$O$_{7}$ {\it Hyperfine Interactions} {\bf 156/157} 103

\bibitem{Beeman} Beeman D and Pincus P 1968 Nuclear spin-lattice relaxation in magnetic insulators {\it Phys. Rev.} {\bf 166} 359

\bibitem{ChapuisTSO} Chapuis Y {\it et al.} 2007 Ground state of the geometrically frustrated compound Tb$_{2}$Sn$_{2}$O$_{7}$  {\it J. Phys.:Condens. Matter} {\bf 19} 446206

\bibitem{RuleTSO1} Rule K C {\it et al.} 2007 Polarized inelastic neutron scattering of the partially ordered Tb$_{2}$Sn$_{2}$O$_{7}$ {\it Phys. Rev. B} {\bf 76} 212405

\bibitem{MirebeauTSO} Mirebeau I {\it et al.} 2008 Investigation of magnetic fluctuations in Tb$_{2}$Sn$_{2}$O$_{7}$ ordered spin ice by high-resolution energy-resolved neutron scattering {\it Phys. Rev. B} {\bf 78} 174416

\bibitem{RuleTSO2} Rule K C {\it et al.} 2009 Neutron scattering investigations of the partially ordered pyrochlore Tb$_{2}$Sn$_{2}$O$_{7}$ {\it J. Phys.:Condens. Matter} {\bf 21} 489005

\end{thebibliography}
\end{document}